\newcommand {\Mpc}   {\mbox{$ h^{-1}$} Mpc\,}

\newcommand {\MLR}   {\mbox{$ \cal M/L $}--ratio\,}

\documentstyle{l-aa}

\begin{document}

\thesaurus{
       (12.03.3;  
        12.12.1;  
        11.03.1;  
        11.11.1)  
          }


\title{The ESO Nearby Abell Cluster Survey
       \thanks{Based on observations collected at the European Southern
               Observatory (La Silla, Chile)}
      }

\subtitle{II. The Distribution of Velocity Dispersions of Rich Galaxy Clusters
	\thanks{Tables~1$a$ and 1$b$ are also available in electronic form
	        at the CDS via anonymous ftp 130.79.128.5}}

\author{A.~Mazure \inst{1}, P.~Katgert \inst{2}, R.~den~Hartog \inst{2},
        A.~Biviano \inst{2,3}, P.~Dubath \inst{4},
        E.~Escalera \inst{5,6}, P.~Focardi \inst{7},
        D.~Gerbal \inst{3,8}, G.~Giuricin \inst{5,9},
        B.~Jones \inst{10},
        O.~Le~F\`evre \inst{8}, M.~Moles \inst{11},
        J.~Perea \inst{11}, \and G.~Rhee \inst{12}
        }

\institute{Laboratoire d'Astronomie Spatiale, Marseille, France \and
           Sterrewacht Leiden, Netherlands \and
           Institut d'Astrophysique de Paris, France \and
           Observatoire de Gen\`eve, Switzerland \& Lick Observatory, Univ. of
California, Santa Cruz, USA \and
           Dipartimento di Astronomia, Universit\`a di Trieste, Italy \and
           Royal Observatory, Edinburgh, UK \and
           Dipartimento di Astronomia, Universit\`a di Bologna, Italy \and
           DAEC, Observatoire de Paris, Universit\'e Paris 7, CNRS (UA 173),
France \and
	   SISSA, Trieste, Italy \and
           Nordita, Copenhagen, Denmark \and
	   Instituto de Astrof\'{\i}sica de Andaluc\'{\i}a, CSIC, Granada, Spain \and
           Department of Physics, University of Nevada, Las Vegas, USA
           }

\offprints{A.~Mazure}

\date{Received date; accepted date}

\maketitle
\markboth{The ESO Nearby Abell Cluster Survey: II. The Distribution
          of Velocity Dispersions}{}

\begin{abstract}

The ESO Nearby Abell Cluster Survey (the ENACS) has yielded 5634
redshifts for galaxies in the directions of 107 rich, Southern
clusters selected from the ACO catalogue (Abell et al.\ 1989). By
combining these data with another 1000 redshifts from the literature,
of galaxies in 37 clusters, we construct a volume-limited sample of
128 $R_{\rm ACO} \geq 1$ clusters in a solid angle of 2.55 sr
centered on the South Galactic Pole, out to a redshift $z=0.1$. For a
subset of 80 of these clusters we can calculate a reliable velocity
dispersion, based on at least 10 (but very often between 30 and 150)
redshifts.

We deal with the main observational problem that hampers an
unambiguous interpretation of the distribution of cluster velocity
dispersions, namely the contamination by fore- and background
galaxies.  We also discuss in detail the completeness of the cluster
samples for which we derive the distribution of cluster velocity
dispersions. We find that a cluster sample which is complete in terms
of the field-corrected richness count given in the ACO catalogue
gives a result that is essentially identical to that based on a
smaller and more conservative sample which is complete in terms of an
intrinsic richness count that has been corrected for superposition
effects.

We find that the large apparent spread in the relation between
velocity dispersion and richness count (based either on visual
inspection or on machine counts) must be largely intrinsic;
i.e.\ this spread is not primarily
due to measurement uncertainties. One of the consequences of the (very)
broad relation between cluster richness and velocity dispersion is
that all samples of clusters that are defined complete with respect
to richness count are unavoidably biased against low-$\sigma_V$
clusters. For the richness limit of our sample this bias operates
only for velocity dispersions less than $\approx$800 km/sec.

We obtain a statistically reliable distribution of global velocity
dispersions which, for velocity dispersions $\sigma_V \ga 800$ km/s,
is free from systematic errors and biases. Above this value of
$\sigma_V$ our distribution agrees very well with the most recent
determination of the distribution of cluster X-ray temperatures, from
which we conclude that $\beta = \sigma_V^2 \mu m_H/kT_X \approx 1$.

The observed distribution $n(>\sigma_V)$, and especially its
high-$\sigma_V$ tail above $\approx$800 km/s, provides a reliable and
discriminative constraint on cosmological scenarios for the formation
of structure. We stress the need for model predictions that produce
exactly the same information as do the observations, namely
dispersions of line-of-sight velocity of galaxies within the
turn-around radius and inside a cylinder rather than a sphere, for a
sample of model clusters with a richness limit that mimics that of
the sample of observed clusters.

\end{abstract}


\section{Introduction}

The present-day distribution of cluster masses contains information
about important details of the formation of large-scale structure in
the Universe. In principle, the distribution of present cluster
masses constrains the form and amplitude of the spectrum of initial
fluctuations, via the tail of high-amplitude fluctuations from which
the clusters have formed, as well as the cosmological parameters that
influence the formation process. Recently, several authors have
attempted to use either the distribution of cluster mass estimates,
or of gauges of the mass (such as the global velocity dispersion, or
the temperature of the X-ray gas) to constrain parameters in
cosmological scenarios. For example, Frenk et al.\ (1990, FWED
hereafter) have attempted to constrain the bias parameter required by
the CDM scenario through a comparison of their predictions from N-body
simulations with the observed distribution of cluster velocity
dispersions and X-ray temperatures. Subsequently, Henry \& Arnaud
(1991) have used the distribution of cluster X-ray temperatures to
constrain the slope and the amplitude of the spectrum of
fluctuations. More recently, Bahcall \& Cen (1992), Biviano et al. (1993),
and White et al.\
(1993) have used the distribution of estimated masses to constrain
the cosmological density parameter, the power-spectrum index,
as well as the bias parameter.

For constraining the {\em slope} of the spectrum of initial
fluctuations through the slope of the mass distribution, unbiased
estimates of the mass (or of a relevant mass gauge) are required. The
latter always require assumptions about either the shape of the
galaxy orbits, the shape of the mass distribution, or about the
distribution of the gas temperature. Therefore gauges of the total
mass that are based on directly observable parameters, such as global
velocity dispersions or central X-ray temperatures, are sometimes
preferable. However, the use of such mass gauges also requires a lot
of care. Global velocity dispersions, although fairly easily
obtained, can be affected by projection effects and contamination by
field galaxies, as discussed by e.g.\ FWED. In addition, velocity
dispersions may depend on the size of the aperture within which they
are determined, because the dispersion of the line-of-sight
velocities often varies with distance from the cluster centre (e.g.\
den~Hartog \& Katgert 1995).

More fundamentally, the velocity dispersion of the galaxies may be a
biased estimator of the cluster potential (or mass) as a result of
dynamical friction and other relaxation processes. In principle, the
determination of the X-ray temperatures is more straightforward.
However, temperature estimates may be affected by cooling flows,
small-scale inhomogeneities (Walsh \& Miralda-Escud\'e 1995), bulk
motions or galactic winds (Metzler \& Evrard 1994). Also,
temperature estimates of high accuracy require high spectral
resolution and are therefore less easy to obtain.

To obtain useful constraints on the {\em amplitude} of the
fluctuation spectrum, it is essential that the completeness of the
cluster sample in the chosen volume is accurately known. The
completeness of cluster samples constructed from galaxy catalogues
obtained with automatic scanning machines, such as the COSMOS and APM
machines (see e.g.\ Lumsden et al.\ 1992, LNCG hereafter, and Dalton
et al.\ 1992) is, in principle, easier to discuss
than that of the ACO catalogue, which until
recently was the only source of cluster samples.  In theory, one is
primarily interested in the completeness with respect to a
well-defined limit in mass. In practice, cluster samples based on
optical catalogues can be defined only with respect to richness, and
the relation between richness and mass seems to be very broad. A further
complication is that all optical cluster catalogues suffer from
superposition effects, which can only be resolved through extensive
spectroscopy. Cluster samples based on X-ray surveys do not suffer
from superposition effects, but they are (of necessity) flux-limited,
and the extraction of volume-limited samples with well-defined
luminosity limits requires follow-up spectroscopy (e.g.  Pierre et
al.\ 1994). The large spread in the relation between X-ray luminosity
and X-ray temperature (e.g.\ Edge \& Stewart 1991) implies that, as
with the optical samples, the construction of cluster samples with a
well-defined mass limit from X-ray surveys is not at all trivial.

In this paper we discuss the distribution of velocity dispersions,
for a volume-limited sample of rich ACO clusters with known
completeness. The discussion is based on the results of our ESO
Nearby Abell Cluster Survey (ENACS, Katgert et al.\ 1995, hereafter
Paper I), which has yielded 5634 reliable galaxy redshifts in the
direction of 107 rich, nearby ACO clusters with redshifts out to
about 0.1. We have supplemented our data with about 1000 redshifts
from the literature for galaxies in 37 clusters.

In Section~2 we describe the construction of a volume-limited sample
of rich clusters. In Section~3 we discuss superposition effects, and
introduce a {\em 3-dimensional} richness (derived from Abell's
projected, 2-dimensional richness). In Section~4 we discuss the
completeness of the cluster sample, and we estimate the spatial
density of rich clusters. In Section~5 we summarize the procedure
that we used for eliminating interlopers, which is essential for
obtaining unbiased estimates of velocity dispersion. In Section~6 we
derive the properly normalized distribution of velocity dispersions.
In Section~7 we compare our distribution with earlier results from
the literature, which include both distributions of velocity
dispersions and of X-ray temperatures. Finally, we also compare our
result with some published predictions from N-body simulations.

\section{The Cluster Sample}

\subsection{Requirements}

The observational determination of the distribution of cluster
velocity dispersions, $n(\sigma_V)$, requires a cluster sample that
is either not biased with respect to $\sigma_V$, or that has a bias
which is sufficiently well-known that it can be corrected for in the
observations or be accounted for in the predictions. If this
condition is fulfilled for a certain range of velocity dispersions,
the {\em shape} of the distribution can be determined over that
range. A determination of the {\em amplitude} of the distribution
requires that the spatial completeness of the cluster sample is also
known.

Until complete galaxy redshift surveys over large solid angles and
out to sufficiently high redshifts become available, it will not be
possible to construct cluster samples that are complete to a
well-defined limit in velocity dispersion. The only possible manner
in which this ideal can at present be approached is by selecting
cluster samples from catalogues that are based on overdensities in
projected distributions of galaxies (or in X-ray surface brightness).
By selecting only the clusters with an apparent richness (i.e.\
surface density in a well-defined range of absolute magnitudes) above
a certain lower limit, one may hope to obtain an approximate lower
limit in intrinsic richness (i.e.\ with fore- and background-galaxies
removed).

By virtue of the general (but very broad) correlation between
richness and velocity dispersion one can then expect to achieve
completeness with respect to velocity dispersion above a lower limit
in $\sigma_V$. Below that limit the cluster sample will be inevitably
incomplete with respect to velocity dispersion, in a manner that is
specific for the adopted richness limit. In other words: observed and
predicted velocity dispersion distributions can be compared directly
above the limiting $\sigma_V$ set by the richness limit. For smaller
values of the velocity dispersion the prediction should take into
account the bias introduced by the particular richness limit.

\subsection{The Southern ACO $R\geq 1$ Cluster Sample}

The ENACS was designed to establish, in combination with data already
available in the literature, a database for a complete sample of
$R\geq 1$ ACO clusters, out to a redshift of $z=0.1$, in a solid angle
of 2.55 sr around the SGP, defined by $b\leq -30\degr$ and $-70\degr
\le \delta \leq 0\degr$ (a volume we will refer to as the `cone'). For
our sample we selected clusters which at the time either had a known
spectroscopic redshift $z \leq 0.1$, or which had $m_{10}\leq
16.9$. Judging from the $m_{10}-z$ relation the clusters with
$m_{10}\leq 16.9$ should include most of the $z\leq 0.1$
clusters. With this selection, the contamination from $z>0.1$ clusters
would clearly be non-negligible due to the spread in the $m_{10}-z$
relation.

\begin{table}
\caption{{\bf a.} The statistical sample of 128 ACO clusters that
         are in the 'cone' and have a main system with $z < 0.1$,
         and 14 relevant secondary systems with $z < 0.1$.}
\begin{flushleft}
\begin{tabular}{lrcrrrr}
\hline
\noalign{\smallskip}
 ACO & $N_{\rm mem}$ & z & $\sigma_V^c$ &
 $C_{\rm ACO}$ & $C_{\rm bck}$ & $C_{\rm 3D}$ \\
\noalign{\smallskip}
\hline
\noalign{\smallskip}
A0013            &  37 & 0.0943 &  886 &  96 &  20.7 &  98.2 \\
A0074$^\ddagger$ &   5 & 0.0654 &    - &  50 &  15.0 &     - \\
A0085$^\ddagger$ & 116 & 0.0556 &  853 &  59 &  15.9 &  55.9 \\
                 &  17 & 0.0779 &  462 &  59 &  15.9 &   7.7 \\
A0087            &  27 & 0.0550 &  859 &  50 &  24.3 &  47.8 \\
A0119$^\dagger$  & 125 & 0.0442 &  740 &  69 &  17.5 &  76.1 \\
A0126$^\ddagger$ &   1 & 0.0850 &    - &  51 &  24.3 &     - \\
A0133$^\ddagger$ &   9 & 0.0566 &    - &  60 &  15.8 &     - \\
A0151$^\dagger$  &  63 & 0.0533 &  669 &  72 &  17.5 &  39.1 \\
                 &  40 & 0.0997 &  857 &  72 &  17.5 &  25.2 \\
                 &  29 & 0.0410 &  395 &  72 &  17.5 &  18.3 \\
A0168            &  76 & 0.0450 &  517 &  89 &  16.5 &  80.2 \\
A0261$^\ddagger$ &   1 & 0.0467 &    - &  63 &  30.5 &     - \\
A0277$^\ddagger$ &   2 & 0.0927 &    - &  50 &  16.1 &     - \\
A0295            &  30 & 0.0426 &  297 &  51 &  24.3 &  52.6 \\
A0303            &   4 & 0.0595 &    - &  50 &  24.3 &  18.6 \\
A0367            &  27 & 0.0907 &  963 & 101 &  19.5 & 108.4 \\
A0415$^\ddagger$ &   1 & 0.0788 &    - &  67 &  20.8 &     - \\
A0420            &  19 & 0.0858 &  514 &  55 &  26.3 &  46.8 \\
A0423$^\ddagger$ &   2 & 0.0795 &    - &  89 &  24.3 &     - \\
A0484$^\ddagger$ &   4 & 0.0386 &    - &  50 &  27.3 &     - \\
A0496$^\ddagger$ & 134 & 0.0328 &  682 &  50 &  16.8 &  57.0 \\
A0500$^\ddagger$ &   1 & 0.0666 &    - &  58 &  17.2 &     - \\
A0514            &  82 & 0.0713 &  874 &  78 &  14.5 &  68.4 \\
A0524            &  26 & 0.0779 &  822 &  74 &  34.4 &  65.6 \\
                 &  10 & 0.0561 &  211 &  74 &  34.4 &  25.3 \\
A2361            &  24 & 0.0608 &  329 &  69 &  27.3 &  70.0 \\
A2362            &  17 & 0.0608 &  340 &  50 &  25.3 &  47.4 \\
A2377$^\ddagger$ &   1 & 0.0808 &    - &  94 &  27.3 &     - \\
A2382$^\ddagger$ &   1 & 0.0648 &    - &  50 &  17.7 &     - \\
A2384$^\ddagger$ &   1 & 0.0943 &    - &  72 &  24.8 &     - \\
A2399$^\ddagger$ &   1 & 0.0587 &    - &  52 &  16.1 &     - \\
A2400$^\ddagger$ &   1 & 0.0881 &    - &  56 &  23.1 &     - \\
A2401            &  23 & 0.0571 &  472 &  66 &  18.6 &  64.9 \\
A2410$^\ddagger$ &   1 & 0.0806 &    - &  54 &  17.7 &     - \\
A2420$^\ddagger$ &   1 & 0.0838 &    - &  88 &  26.3 &     - \\
A2426            &  15 & 0.0978 &  846 & 114 &  26.3 &  58.5 \\
                 &  11 & 0.0879 &  313 & 114 &  26.3 &  42.9 \\
A2436            &  14 & 0.0914 &  530 &  56 &  27.3 &  61.4 \\
A2480            &  11 & 0.0719 &  862 & 108 &  23.0 &  80.0 \\
A2492$^\ddagger$ &   2 & 0.0711 &    - &  62 &  18.6 &     - \\
A2500            &  13 & 0.0895 &  477 &  71 &  82.2 &  55.3 \\
                 &  12 & 0.0783 &  283 &  71 &  82.2 &  51.0 \\
A2502$^\ddagger$ &   0 & 0.0972 &    - &  58 &  24.3 &     - \\
A2528$^\ddagger$ &   1 & 0.0955 &    - &  73 &  22.0 &     - \\
A2538$^\ddagger$ &  42 & 0.0832 &  861 &  83 &  19.1 &  95.3 \\
A2556$^\ddagger$ &   2 & 0.0865 &    - &  67 &  21.2 &     - \\
A2559$^\ddagger$ &   1 & 0.0796 &    - &  73 &  28.3 &     - \\
A2566$^\ddagger$ &   1 & 0.0821 &    - &  67 &  28.4 &     - \\
A2569            &  36 & 0.0809 &  481 &  56 &  24.3 &  70.5 \\
A2599$^\ddagger$ &   1 & 0.0880 &    - &  84 &  16.2 &     - \\
A2638$^\ddagger$ &   1 & 0.0825 &    - & 123 &  30.5 &     - \\
A2644            &  12 & 0.0688 &  259 &  59 &  24.3 &  28.6 \\
\noalign{\smallskip}
\hline
\end{tabular}
\end{flushleft}
\end{table}

\begin{table}
\addtocounter{table}{-1}
\caption[]{{\bf a.} $-$ continued $-$}
\begin{flushleft}
\begin{tabular}{lrcrrrr}
\hline
\noalign{\smallskip}
 ACO & $N_{\rm mem}$ & z & $\sigma_V^c$ &
 $C_{\rm ACO}$ & $C_{\rm bck}$ & $C_{\rm 3D}$ \\
\noalign{\smallskip}
\hline
\noalign{\smallskip}
A2670$^\ddagger$ & 219 & 0.0762 &  908 & 142 &  15.9 & 114.1 \\
A2717$^\dagger$  &  56 & 0.0490 &  512 &  52 &  10.8 &  43.4 \\
A2734            &  77 & 0.0617 &  581 &  58 &  12.3 &  45.9 \\
A2755            &  22 & 0.0949 &  789 & 120 &  28.2 &  90.6 \\
A2764            &  19 & 0.0711 &  788 &  55 &  19.6 &  59.1 \\
A2765            &  16 & 0.0801 &  905 &  55 &  47.7 &  58.7 \\
A2799            &  36 & 0.0633 &  424 &  63 &  20.2 &  71.3 \\
A2800            &  34 & 0.0636 &  430 &  59 &  18.6 &  57.4 \\
A2819            &  50 & 0.0747 &  406 &  90 &  23.8 &  45.2 \\
                 &  44 & 0.0867 &  359 &  90 &  23.8 &  39.7 \\
A2854            &  22 & 0.0613 &  369 &  64 &  28.1 &  58.0 \\
A2889$^\ddagger$ &   1 & 0.0667 &    - &  65 &  22.0 &     - \\
A2911            &  31 & 0.0808 &  576 &  72 &  21.3 &  65.7 \\
A2915            &   4 & 0.0864 &    - &  55 &  25.0 &     - \\
A2923            &  16 & 0.0715 &  339 &  50 &  42.3 &  44.8 \\
A2933            &   9 & 0.0925 &    - &  77 &  28.2 &  86.1 \\
A2954            &   6 & 0.0566 &    - & 121 &  32.2 &  38.3 \\
A2955$^\ddagger$ &   0 & 0.0989 &    - &  56 &  34.4 &     - \\
A3009            &  12 & 0.0653 &  514 &  54 &  21.3 &  56.5 \\
A3040$^\ddagger$ &   1 & 0.0923 &    - &  69 &  17.2 &     - \\
A3093            &  22 & 0.0830 &  435 &  93 &  22.5 &  63.5 \\
A3094            &  66 & 0.0672 &  653 &  80 &  18.6 &  65.8 \\
A3107$^\ddagger$ &   0 & 0.0875 &    - &  61 &  20.4 &     - \\
A3108            &   7 & 0.0625 &    - &  73 &  30.5 &  51.7 \\
                 &   5 & 0.0819 &    - &  73 &  30.5 &  36.9 \\
A3111            &  35 & 0.0775 &  770 &  54 &  18.6 &  52.9 \\
A3112            &  67 & 0.0750 &  950 & 116 &  25.2 &  92.8 \\
A3122            &  87 & 0.0643 &  755 & 100 &  21.4 &  88.7 \\
A3126$^\ddagger$ &  38 & 0.0856 & 1041 &  75 &  29.3 &  88.1 \\
A3128$^\dagger$  & 180 & 0.0599 &  802 & 140 &  19.4 & 129.3 \\
                 &  12 & 0.0395 &  386 & 140 &  19.4 &   8.6 \\
                 &  12 & 0.0771 &  103 & 140 &  19.4 &   8.6 \\
A3135$^\ddagger$ &   1 & 0.0633 &    - & 111 &  28.6 &     - \\
A3144            &   1 & 0.0423 &    - &  54 &  16.3 &     - \\
A3151            &  34 & 0.0676 &  747 &  52 &  23.8 &  60.0 \\
A3152$^\ddagger$ &   0 & 0.0891 &    - &  51 &  26.0 &     - \\
A3153$^\ddagger$ &   0 & 0.0958 &    - &  64 &  26.0 &     - \\
A3158            & 105 & 0.0591 & 1005 &  85 &  10.8 &  82.5 \\
A3194            &  32 & 0.0974 &  790 &  83 &  13.3 &  93.5 \\
A3202            &  27 & 0.0693 &  433 &  65 &  28.1 &  61.3 \\
A3223            &  68 & 0.0601 &  636 & 100 &  14.2 &  69.6 \\
A3264            &   5 & 0.0978 &    - &  53 &  37.6 &  41.2 \\
A3266$^\ddagger$ & 158 & 0.0589 & 1105 &  91 &  19.0 &  97.1 \\
A3301            &   5 & 0.0536 &    - & 172 &   7.3 &     - \\
A3330$^\ddagger$ &   1 & 0.0910 &    - &  52 &  18.0 &     - \\
A3334$^\ddagger$ &  32 & 0.0965 &  671 &  82 &  29.3 &  86.9 \\
A3341            &  63 & 0.0378 &  566 &  87 &  23.9 &  59.2 \\
                 &  15 & 0.0776 &  751 &  87 &  23.9 &  14.1 \\
A3351$^\ddagger$ &   0 & 0.0819 &    - & 114 &  35.0 &     - \\
A3360$^\ddagger$ &  36 & 0.0848 &  801 &  85 &  34.6 & 107.7 \\
A3651            &  78 & 0.0599 &  661 &  75 &  33.2 &  91.8 \\
A3667$^\dagger$  & 162 & 0.0556 & 1059 &  85 &  33.2 &  85.1 \\
A3677            &   8 & 0.0912 &    - &  60 &  25.0 &  37.7 \\
A3682            &  10 & 0.0921 &  863 &  66 &  41.1 &  97.3 \\
A3691            &  33 & 0.0873 &  792 & 115 &  40.9 & 142.9 \\
\noalign{\smallskip}
\hline
\end{tabular}
\end{flushleft}
\end{table}

\begin{table}
\addtocounter{table}{-1}
\caption[]{{\bf a.} $-$ continued $-$}
\begin{flushleft}
\begin{tabular}{lrcrrrr}
\hline
\noalign{\smallskip}
 ACO & $N_{\rm mem}$ & z & $\sigma_V^c$ &
 $C_{\rm ACO}$ & $C_{\rm bck}$ & $C_{\rm 3D}$ \\
\noalign{\smallskip}
\hline
\noalign{\smallskip}
A3693            &  16 & 0.0910 &  585 &  77 &  25.0 &  49.5 \\
A3695            &  81 & 0.0893 &  845 & 123 &  39.5 & 137.2 \\
A3696            &  12 & 0.0882 &  428 &  58 &  30.6 &  88.6 \\
A3698$^\ddagger$ &   1 & 0.0198 &    - &  71 &   7.7 &     - \\
A3703            &  18 & 0.0735 &  455 &  52 &  27.3 &  44.6 \\
                 &  13 & 0.0914 &  697 &  52 &  27.3 &  32.2 \\
A3705            &  29 & 0.0898 & 1057 & 100 &  32.0 &  93.3 \\
A3716$^\ddagger$ &  65 & 0.0448 &  781 &  66 &  11.6 &  61.6 \\
A3733            &  41 & 0.0389 &  696 &  59 &   4.7 &  59.4 \\
A3744            &  66 & 0.0381 &  559 &  70 &  10.5 &  62.8 \\
A3764            &  38 & 0.0757 &  671 &  53 &  24.0 &  68.1 \\
A3781            &   4 & 0.0571 &    - &  79 &  16.2 &  25.4 \\
                 &   4 & 0.0729 &    - &  79 &  16.2 &  25.4 \\
A3795            &  13 & 0.0890 &  336 &  51 &  31.9 &  77.0 \\
A3799            &  10 & 0.0453 &  428 &  50 &  24.9 &  50.0 \\
A3806            &  84 & 0.0765 &  813 & 115 &  11.7 &  89.4 \\
A3809            &  89 & 0.0620 &  499 &  73 &  20.8 &  55.5 \\
A3822            &  84 & 0.0759 &  969 & 113 &  22.5 & 112.8 \\
A3825            &  59 & 0.0751 &  698 &  77 &  12.0 &  58.4 \\
A3826$^\ddagger$ &   1 & 0.0754 &    - &  62 &  13.2 &     - \\
A3827            &  20 & 0.0984 & 1114 & 100 &  28.4 & 116.7 \\
A3844$^\ddagger$ &   1 & 0.0730 &    - &  52 &  23.0 &     - \\
A3879            &  46 & 0.0669 &  516 & 114 &  39.5 &  85.0 \\
A3897            &  10 & 0.0733 &  548 &  63 &  21.3 &  64.8 \\
A3911$^\ddagger$ &   1 & 0.0960 &    - &  58 &  30.5 &     - \\
A3921            &  32 & 0.0936 &  585 &  93 &  25.0 &  99.3 \\
A3969$^\ddagger$ &   1 & 0.0699 &    - &  55 &  40.5 &     - \\
A4008            &  27 & 0.0549 &  424 &  66 &  36.0 &  64.1 \\
A4010            &  30 & 0.0957 &  615 &  67 &  28.2 &  79.3 \\
A4038$^\ddagger$ &  51 & 0.0292 &  839 & 117 &  17.1 & 110.4 \\
A4053            &  17 & 0.0720 &  731 &  64 &  16.2 &  43.9 \\
                 &   9 & 0.0501 &    - &  64 &  16.2 &  23.2 \\
A4059$^\ddagger$ &  10 & 0.0488 &  526 &  66 &  11.0 &  69.9 \\
A4067$^\ddagger$ &  30 & 0.0989 &  719 &  72 &  30.5 &  75.0 \\
\noalign{\smallskip}
\hline
\noalign{\smallskip}
\end{tabular}
\end{flushleft}
{{\bf Notes:}
col.(1): A dagger indicates a combination of data from the ENACS and from
the literature, a double dagger indicates that only data from the
literature were used;
col.(2): secondary systems are listed if they contain either $\geq$10
redshifts or $\geq$50\% of the number of redshifts of the
main system;
col.(3): redshift values (or photometric estimates, indicated by $N_{\rm
z}$=0) of clusters for which no ENACS data exist, were taken
from Abell et al.\ (1989), Struble \& Rood (1991), Dalton et al.\ (1994),
Postman et al.\ (1992), West (private communication) and Quintana \&
Ram\'{\i}rez (1995)}
\end{table}

At present, after completion of our project and with other new data in
the literature, the region defined above contains 128 $R\geq 1$ ACO
clusters with a measured or estimated redshift $z\leq 0.1$. A
spectroscopically confirmed redshift $z\leq 0.1$ is available for 122
clusters, while for the remaining 6 a redshift $\leq 0.1$ has been
estimated on the basis of photometry. The redshift values (or
estimates), if not from the ENACS, were taken from Abell et al.\
(1989), Struble \& Rood (1991), Peacock and West (1992, and private
communication), Postman et al.\ (1992), Dalton et al.\ (1994) and
Quintana \& Ram\'{\i}rez (1995).

We will show below that the 128 clusters form a sample that can be
used for statistical analysis. Of the 122 redshift surveys of
clusters with $z \leq 0.1$ in the specified region, 78 were
contributed to by the ENACS, either
exclusively or in large measure. In 80 of the 122 redshift surveys we
find at least one system with 10 or more member galaxies. Of the
latter 80 surveys, 68 were contributed to by our survey.

In Tab.~1a we list several properties of the main systems in the
direction of the 128 clusters (in the `cone' and with $z < 0.1$) that
constitute the sample on which we will base our discussion of the
distribution of cluster velocity dispersions, as well as the
properties of 14 subsystems with $z < 0.1$ that either have 10
redshifts or at least half the number of redshifts in the main
system. In Tab.~1b we list the same type of data for the other systems
described in Paper I, with at least 10 members so that a velocity
dispersion could be determined. As the latter are outside the `cone'
defined above (or have $z > 0.1$), they have not been used in the
present discussion but could be useful for other purposes. A
description of the ways in which the data in Tabs. 1$a$ and 1$b$ have
been obtained will be given in the next Sections.

\begin{table}
\addtocounter{table}{-1}
\caption{{\bf b.} Main systems in the `cone' with $z > 0.1$ and $N
                  \ge 10$ (with relevant secondary systems), and systems
                  outside the `cone' with $N \ge 10$.}

\begin{flushleft}
\begin{tabular}{lrcrrrr}
\hline
\noalign{\smallskip}
 ACO & $N_{\rm mem}$ & z & $\sigma_V^c$ &
 $C_{\rm ACO}$ & $C_{\rm bck}$ & $C_{\rm 3D}$ \\
\noalign{\smallskip}
\hline
\noalign{\smallskip}
A0118           & 30 & 0.1148 & 649 & 77 & 35.8 & 89.1 \\
A0229           & 32 & 0.1139 & 856 & 77 & 24.3 & 83.1 \\
A0380           & 25 & 0.1337 & 703 & 82 & 35.8 & 71.8 \\
A0543           & 10 & 0.0850 & 413 & 90 &244.2 &139.3 \\
A0548$^\dagger$ &323 & 0.0416 & 842 & 92 & 12.2 & 88.4 \\
                & 21 & 0.1009 & 406 &    &      &  5.4 \\
                & 15 & 0.0868 &1060 &    &      &  3.9 \\
A0754$^\dagger$ & 90 & 0.0543 & 749 & 92 & 17.0 &101.6 \\
A0957           & 34 & 0.0447 & 741 & 55 & 16.8 & 67.8 \\
A0978           & 56 & 0.0544 & 498 & 55 & 16.1 & 61.4 \\
A1069           & 35 & 0.0650 &1120 & 45 & 17.2 & 54.5 \\
A1809$^\dagger$ & 58 & 0.0791 & 702 & 78 & 16.0 & 81.7 \\
A2040           & 37 & 0.0461 & 673 & 52 & 15.9 & 58.4 \\
A2048           & 25 & 0.0972 & 668 & 75 & 17.7 & 59.4 \\
A2052$^\dagger$ & 62 & 0.0350 & 655 & 41 & 17.5 & 52.5 \\
A2353           & 24 & 0.1210 & 599 & 51 & 26.3 & 59.8 \\
A2715$^*$       & 14 & 0.1139 & 556 &112 & 30.5 & 58.7 \\
A2778           & 17 & 0.1018 & 947 & 51 & 11.9 & 26.7 \\
                & 10 & 0.1182 & 557 &    &      & 15.7 \\
A2871           & 18 & 0.1219 & 930 & 92 & 48.0 & 63.0 \\
                & 14 & 0.1132 & 319 &    &      & 49.0 \\
A3141           & 15 & 0.1058 & 646 & 55 & 16.2 & 48.5 \\
A3142$^*$       & 21 & 0.1030 & 814 & 78 & 13.0 & 50.3 \\
                & 12 & 0.0658 & 785 &    &      & 28.7 \\
A3354           & 57 & 0.0584 & 383 & 54 &  9.8 & 33.6 \\
A3365           & 32 & 0.0926 &1153 & 68 & 32.0 & 91.5 \\
A3528           & 28 & 0.0526 & 969 & 70 &  6.2 & 54.7 \\
A3558$^\dagger$ &328 & 0.0479 & 939 &226 &  8.7 &127.0 \\
A3559           & 39 & 0.0461 & 443 &141 &  9.3 & 85.0 \\
                & 11 & 0.1119 & 537 &    &      & 24.0 \\
A3562           &114 & 0.0490 &1048 &129 & 12.6 &140.4 \\
A3864           & 32 & 0.1033 & 940 & 60 & 29.5 & 69.8 \\
\noalign{\smallskip}
\hline
\noalign{\smallskip}
\end{tabular}
\end{flushleft}
{{\bf Notes:}
col.(1): an asterisk indicates that the system is in the `cone', a dagger
indicates a combination of data from the ENACS and from the literature;
col.(2): secondary systems are listed if they contain either $\geq$10
redshifts or $\geq$50\% of the number of redshifts of the
main system}
\end{table}

\section{Superposition Effects in the ACO Cluster Catalogue}

\begin{figure*}[htb]
\vbox{}
\caption[]{\\
 $a)$ The distribution of $f_{\rm main}$, the fraction of galaxies in
      the largest system (determined from redshift surveys), for the
      80 systems with $N\geq 10$ and $z\leq 0.1$.
 $b)$ The relation between $f_{\rm ACO}$
      (= $C_{\rm ACO}/(C_{\rm ACO} + C_{\rm bck})$) and $f_{\rm main}$.
 $c)$ The relation between $C_{\rm ACO}$ and
      $C_{\rm 3D}\,=f_{\rm main}\times (C_{\rm ACO}+C_{\rm bck})$;
      the horizontal dashed line indicates $C_{\rm ACO}=50$,
      the vertical dashed line $C_{\rm 3D}=50$.}
\end{figure*}

It is clear from the redshift distributions towards our target
clusters in the ENACS (see Fig.~7 in Paper
I) that for most clusters in the $R_{\rm ACO} \geq 1$ sample the
fraction of fore- and background galaxies is non-negligible.

In Paper I we have discussed how one can identify the fore- and
background galaxies, namely as the `complement' of the galaxies in
the physically relevant systems. In order to identify the latter, we
used a fixed velocity gap to decide whether two galaxies in the
survey that are adjacent in redshift, are part of the same system or
of two separate systems. The minimum velocity difference that defines
galaxies to be in separate systems was determined for the ENACS from the
sum of the 107 distributions of the velocity gap between galaxies
that are adjacent in velocity. We found that a gap-width of 1000 km/s
is sufficient to identify systems, that it does not break up systems
inadvertently, and is conservative in the sense that it does not
eliminate outlying galaxies of a system.  Note that the gap-size of
1000 km/s is geared to the sampling in our survey and to the average
properties of the redshift systems; for other datasets the required
gap-size may be different. The systems that result from applying this
procedure to the ENACS data are given in Tab.~6 of Paper I.

Having identified the systems in the redshift surveys of our
clusters we can quantify the effect of the superposition of fore- and
background galaxies on the ACO richness estimates. Our observing
strategy has been to obtain, for the target clusters, redshifts for
the $N$ brightest galaxies in a field consisting of 1 to 3 circular
apertures with a diameter of $\approx0.5\degr$. For most clusters in our
programme this corresponds roughly to the size of the field in which
the richness count was determined. We can therefore estimate an
intrinsic `3-D' richness of a cluster as the product of the fraction
$f_{\rm main}$ of galaxies that reside in the main system (i.e.\ in
the system with the largest number of members) with the {\em total}
galaxy count obtained by Abell et al.\ (1989).

The total count is not available in the ACO catalogue and must be
recovered as the sum of the corrected count $C_{\rm ACO}$ published
by Abell et al.\ (1989) and the correction for the contribution of
the field, $C_{\rm bck}$, that they subtracted from their measured
total count. The intrinsic 3-D richness thus follows as $C_{\rm 3D} =
f_{\rm main} \times (C_{\rm ACO} + C_{\rm bck})$, in which we replace
the statistical field corrections of Abell et al. (1989) (based on
integrals of the galaxy luminosity function) by field corrections
based on redshift surveys.

The first ingredient in the calculation of the intrinsic 3~-~D richness
is $f_{\rm main}$. In Fig.~1$a$ we show the distribution of the
fraction $f_{\rm main}$ for the 80 redshift surveys with $N\geq 10$
and $z\le 0.1$ in our sample. We estimated $f_{\rm main}$ only for
systems with at least 10 measured redshifts, because for $N<10$ the
definition of systems is not very stable, so that the
determination of $f_{\rm main}$ is likewise not very reliable.
According to Girardi et al.\ (1993), the minimum number of galaxies
required to obtain a reliable estimate of $\sigma_V$ also happens to
be about 10. We find that, on average, $\approx$ 73\% of the galaxies
in our redshift surveys towards $R_{\rm ACO} \geq 1$ clusters with
$z\la 0.1$ belongs to the main system.

The correction $C_{\rm bck}$, which accounts for the contribution of
the field to the total count, was derived as follows. Abell et al.\
(1989) describe a parametrization of the background correction, which
is the number of field galaxies down to a limiting magnitude of
$m_3+2$ (the limit of the uncorrected richness count), in the same
aperture in which the total count was made. The latter has a diameter
that is based on the estimated distance through the $m_{10}-z$
relation.  To calculate $C_{\rm bck}$ for a cluster, it is thus
necessary to have its $m_3$ and $m_{10}$. These parameters are known
for those 65 of the 80 clusters in our sample that are in the southern
part of the ACO catalogue. The other 15 clusters were described by
Abell (1958), who did not use a parametrized estimate of the
background, nor did he list $m_3$. Instead, he estimated $C_{\rm bck}$
by counting all galaxies down to $m_3+2$ in a field near each cluster
that clearly did not contain another cluster. To recover an
approximate value of $C_{\rm bck}$ for these 15 clusters we first
estimated $m_3$ from $m_{10}$. Using 97 $R\ge 1$ clusters in our
sample out to $z=0.1$ we found $m_3 = 0.987\, m_{10} - 0.608$, with a
spread of 0.30 mag. Finally, we used the parametrization of Abell et
al.\ (1989) for these 15 clusters to calculate $C_{\rm bck}$ (which
should be very close but need not identical to the value subtracted
by Abell).

In Fig.~1$b$ we show the relation between $f_{\rm main}$ and $f_{\rm
ACO} (= C_{\rm ACO} / (C_{\rm ACO} + C_{\rm bck})$. The average and
median values of $f_{\rm ACO}$ are both 0.76, i.e.\ practically
identical to the corresponding values for $f_{\rm main}$. So, {\em on
average}, the field correction $C_{\rm bck}$ applied by Abell et al.
(1989) was almost the same as the field correction we derive from our
redshift surveys. However, $f_{\rm main}$ spans a much wider range
than does $f_{\rm ACO}$. It thus appears that the field correction
of Abell et al.\ (1989) has probably introduced a considerable noise
in the field-corrected richness estimates. The reason for this is
that their correction was based on an `average field', while for an
individual cluster the actual field may differ greatly from the
average.

This conclusion is supported by the data in Fig.~1$c$, where we show
the relation between $C_{\rm ACO}$, the count corrected for the model
field contribution according to Abell et al. (1989), and $C_{\rm 3D}$, the
intrinsic 3-D count calculated using $f_{\rm main}$, which thus takes
into account the actual field contamination for each cluster
individually. Statistically, $C_{\rm ACO}$ and $C_{\rm 3D}$ appear to
measure the same quantity, i.e. the field correction of Abell et al. (1989)
is, {\em on average}, in very good agreement with our estimates from the
redshift surveys. However, the variations in the real field with respect to
the average field must be mainly responsible for the very large spread
in the values of $C_{\rm 3D}$ for a fixed value of $C_{\rm ACO}$.
As the distribution of points in Fig.1~$c$ seems to be very
symmetric around the $C_{\rm ACO} = C_{\rm 3D}$ -line, we will later
assume that the statistical properties of a complete cluster sample
with $C_{\rm ACO} \ge 50$ are not different from those of a sample
with $C_{\rm 3D} \ge 50$.

\begin{figure*}[htb]
\vbox{}
\caption[]{\\
 $a)$ The dependence of $f_{\rm main}$ (the fraction of galaxies in the
      largest system) on the redshift of the main system.
 $b)$ The dependence of intrinsic richness count $C_{\rm 3D}$ on
      the redshift of the system.
 $c)$ The dependence of $C_{\rm ACO}+C_{\rm bck}$ on $f_{\rm main}$.}
\end{figure*}

For an individual cluster, $C_{\rm 3D}$ is obviously a much more
meaningful parameter than $C_{\rm ACO}$. Yet, one has to be aware of
possible systematic effects that may affect its use. First,
as $C_{\rm 3D}$ involves $f_{\rm main}$ any bias in the determination
of $f_{\rm main}$ could also enter $C_{\rm 3D}$. The number of
unrelated fore- and background galaxies is likely to depend on the
redshift of a cluster, and therefore $f_{\rm main}$ might depend on
redshift.  However, as is evident from Fig.~2$a$, there is hardly any
indication in our data that this is the case. At most, there may be a
tendency for a slight bias against low values of $f_{\rm main}$ at
the lowest redshifts. This is consistent with the fact that for the
nearest clusters the field contribution is low and may not be very
easy to determine properly. In principle, a slight bias against low
values of $f_{\rm main}$ could result in a slight bias against low
values of $C_{\rm 3D}$ for nearby clusters. But, as can be seen from
Fig.~2$b$, there is no indication that for nearby clusters the
$C_{\rm 3D}$ values are higher than average.

Secondly, there is a general tendency to select preferentially the
richer clusters at higher redshifts, and a bias could therefore exist
against the less rich systems at higher redshifts. Although the
systems with the highest values of $C_{\rm 3D}$ are indeed found near
our redshift limit, there is no evidence in Fig.~2$b$ that there is a
strong bias against systems with $C_{\rm 3D} \approx 50$ near the
redshift limit.

Thirdly, the full problem of the superposition of two rich systems
along the line of sight is not appreciated in the simple definition
of $C_{\rm 3D}$, and it is certainly possible that if two $R_{\rm
ACO} \geq 1$ systems are observed in superposition the most distant
one may not be recognized as such. Fortunately, the probability of
such a situation to occur is low. As is clear from Fig.~2$c$, there
is no tendency for clusters with a high total count $C_{\rm ACO} +
C_{\rm bck}$ to have a smaller value of $f_{\rm main}$, as would be
expected if superposition contributed significantly to the richness.
As a matter of fact, given the density of $R\geq 1$ clusters (see
next Section), we expect that for our sample of 128 clusters there is
a probability of about 1\% that a superposition of two $R\geq 1$
clusters occurs in our data. This is consistent with the fact that in
only one case, viz. that of A2500, we observe a secondary system with
a value of $C_{\rm 3D} > 50$.

In principle, the sample of clusters with $C_{\rm 3D}\geq 50$ is to be
preferred over the one with $C_{\rm ACO}\geq 50$, as in the former the
effects of superposition have been accounted for in a proper
way. However, $C_{\rm 3D}$ cannot be used as the main selection
criterion for a cluster sample, because it requires $f_{\rm main}$ to
be available for all clusters. From Fig.~1$c$ it appears that, in a
statistical sense, a cluster sample with $C_{\rm ACO} \ge 50$ can be
used as a substitute for a sample with $C_{\rm 3D}\ga 50$ as, apart
from the large scatter, the two richnesses are statistically
equivalent. As a result of the large scatter in Fig.~1$c$, one can
define a subsample of clusters from the $C_{\rm ACO} \ge 50$ sample
that is complete in terms of $C_{\rm 3D}$ only if one limits the
subsample to systems with $C_{\rm 3D} \ge 75$.

\section{The Completeness of the Sample and the Density of Rich Clusters}

In the following discussion and in the remaining sections of this
paper we will use the term ``cluster'' to refer to the main system in
Tab.~1$a$, i.e. the system with the largest number of redshifts in
each pencil beam, unless we explicitly state otherwise. Hence, the 14
secondary systems in Tab.~1$a$ are not included, nor are the systems
in Tab.1$b$, as the latter are not in the `cone' defined in
Section~2.2, or have $z > 0.1$.

We have estimated the completeness of our cluster sample with respect
to redshift from the distribution of the number of clusters in 10
concentric shells, each with a volume equal to one-tenth of the total
volume out to $z=0.1$. The result is shown in Fig.~3. The dashed line
shows the distribution for all 128 $C_{\rm ACO}\geq 50$ clusters out
to a redshift of 0.1. The solid line represents the subset of 80
clusters with at least 10 redshifts (for which a velocity dispersion
is therefore available). Finally, the dotted line shows the
distribution for the subset of 33 clusters with $N\geq 10$ {\em and}
$C_{\rm 3D}\geq 75$. Note that in Tab.~1a there are 34 clusters with
$C_{\rm 3D}\geq 75$ but one of these, A2933, has only 9 redshifts in
the main system, whereas the total number of redshifts measured was
sufficient to estimate $f_{\rm main}$ and, hence, $C_{\rm 3D}$.

{}From Fig.~3 it appears that the sample of 128 clusters with $C_{\rm
ACO} \ge 50$ has essentially uniform density, except for a possible
($\approx 2\sigma$) `excess' near $z=0.06$, and an apparent `shortage'
of clusters in the outermost bins. The `excess' is at least partly due
to the fact that several of the clusters in the Horologium-Reticulum
and the Pisces-Cetus superclusters are in our cluster sample (see
Paper I).  As we will discuss in detail in the next Section, the
`shortage' towards $z = 0.1$ is probably due to a combination of two
effects. Firstly, some clusters that should have been found by Abell
et al.\ to have $R_{\rm ACO} \ge 1$ and $m_{10}\leq 16.9$ were
not. Secondly, near the redshift limit of our sample Galactic
obscuration may have caused some clusters to be excluded from the
sample that they do belong to.

The subset of 80 clusters for which at least 10 redshifts are
available appears to have essentially uniform density in the inner
half of the volume, but a significantly lower density in the outer
half. This apparent decrease is due to the fact that, for obvious
reasons, the average number of {\em measured} redshifts decreases
with increasing redshift; so much so that for $z\ga 0.08$ the
fraction of clusters with less than 10 redshifts is about 40\%.
Finally, the density of the subset of 33 clusters with $C_{\rm
3D}\geq 75$ appears constant out to a redshift of 0.1. This is
consistent with the fact that none of the selection effects that
operate for the two other samples are expected to affect the richest
clusters.

\subsection{The Sample of 128 Clusters with $R_{\rm ACO} \geq 1$}

In constructing our `complete' sample, we have applied a limit
$m_{10}\le 16.9$ for the cluster candidates without a spectroscopic
redshift. This limit was chosen so that we would include essentially
all clusters with $z < 0.1$. It is possible that a few clusters have
been missed, but it is very difficult to estimate from first
principles how many clusters with $z \le 0.1$ have been missed due to
the $m_{10}$ limit, and we will not try to make a separate estimate
for this effect. However, it is important to realize that the few
clusters that we may have missed as a result of the $m_{10}$ limit
are unlikely to have $z < 0.08$.

\begin{figure}[htb]
\vbox{}
\caption[]{
 The number of clusters in 10 concentric shells, each with a volume
 equal to one-tenth of the total volume out to $z=0.1$. The ordinate
 is volume expressed as a fraction of the total volume out to $z = 0.1$.
 The dashed line represents all 128 clusters in
 the sample, the solid line the 80 clusters with $N\geq 10$, and the
 dotted line the 33 clusters with $C_{\rm 3D}\geq 75$.}
\end{figure}

It is possible that some clusters that should have been included were
either not recognized by Abell (1958) or Abell et al.\ (1989), or
have had their richness underestimated and have thus not made it into
our sample. To a large extent, the magnitude of this effect can be
estimated from a comparison with cluster catalogues based on machine
scanning of plates. Below, we will describe such an estimate. As
with the $m_{10}$ limit, it is likely that clusters that have been
missed for this reason are mostly in the further half of the volume.

Recently, two cluster catalogues that are based on galaxy catalogues
obtained with machine scans of photographic plates have become
available, namely the Edinburgh-Durham Cluster Catalogue (EDCC) by
LNCG, and the APM cluster catalogue by Dalton et
al.\ (1992). We now proceed to estimate, from a comparison with the
EDCC, how many clusters with $R_{\rm ACO} \geq 1$ (or $C_{\rm ACO}
\ge 50$) and $m_{10}\le 16.9$ may have been missed by Abell et al. (1989).

In the following we will assume that the $C_{\rm ACO} \geq 50$
criterion translates into a limit of $C_{\rm EDCC} \geq 30$ in the EDCC
richness count. This assumption is supported by several pieces of
evidence. First, the shift between the distributions of richness
count (see Fig.~3 in LNCG) supports this conclusion, and in
particular the respective richness values at which the incompleteness
sets in. Second, it is also consistent with the apparent offset in
the relation between $C_{\rm ACO}$ and $C_{\rm EDCC}$ (see Fig.~6 in
LNCG). The `offset' between the two richness counts is probably
largely due to different methods used in correcting for the field.
Third, the very large spread in the relation between the two richness
counts, the reason for which is not so obvious, results in about
one-third of the ACO clusters with $C_{\rm ACO}\geq 50$ having a
count $C_{\rm EDCC} < 30$. Conversely, about one-third of the
clusters in the ACO catalogue for which LNCG obtained a count of more
than 30 does not meet the $R_{\rm ACO} \geq 1$ criterion (i.e.\ has
$C_{\rm ACO} < 50$).

One can now try to estimate how many clusters with $R_{\rm ACO} \geq
1$ in our volume have been missed by ACO. Note that the complementary
question, namely how many clusters in the ACO catalogue with $R_{\rm
ACO} \geq 1$ and $m_{10} \leq 16.9$ do not exist according to LNCG,
is not relevant for the present argument, as such ACO cluster
candidates will have been shown by spectroscopy to be non-existent
(there are probably one or two examples in the ENACS). As to the
first question we find, from Fig.~10 in LNCG that of the clusters in
the EDCC without a counterpart in the ACO catalogue, only 5
have $m_{10}(b_J) \leq 17.7$ (which corresponds to $m_{10}(V) \leq
16.9$) {\em as well as} a richness count $C_{\rm EDCC} \geq 30$
(which we assume to correspond to $C_{\rm ACO}\geq 50$). Note that
the EDCC is at high galactic latitude (with $b \la -45\degr$), so
that Galactic obscuration does not play a r\^ole in this comparison.

Among these 5 clusters, there are two for which the richness is
uncertain, but unlikely to be less than 30. We therefore conclude
that these 5 clusters are most likely true $R\geq 1$ clusters
that were missed by ACO (for whatever reason). Two of these 5
clusters have $m_{10}(b_J) \leq 17.1$ while the others have
$m_{10}(b_J) > 17.3$. We assume therefore that 2 clusters with $z \la
0.08$ have been missed in the solid angle of the EDCC by Abell et
al. (1989), and that the other 3 clusters missed have $0.08 \la z \leq
0.1$. As the solid angle of our sample is 5.1 times larger than that
of the EDCC, we estimate that from our sample 10 $R_{\rm ACO} \geq 1$
clusters with $z \la 0.08$, and 15 $R_{\rm ACO} \geq 1$ clusters with
$0.08 \la z \leq 0.1$ are missing.

At first sight it might seem that these numbers should be reduced by
one-third, because of the fact that only two-thirds of the $C_{\rm
EDCC}\geq 30$ clusters have $C_{\rm ACO}\geq 50$. However, that would
ignore the fact that among the clusters with $C_{\rm EDCC} < 30$, a
certain fraction has $C_{\rm ACO}\geq 50$, of which a few are also
likely to have been missed by Abell et al. (1989). On the other hand, we
consider these estimates of the number of clusters missing from our
sample as upper limits, for the following reason. Near the richness
completeness limit of a cluster sample there is some arbitrariness in
accepting and rejecting clusters due to the uncertainties in the
richness estimates. Because the number of clusters increases with
decreasing $C_{\rm ACO}$, it is likely that we have accepted slightly
more clusters than we should have done, as a result of the noise in
the $C_{\rm ACO}$ estimates.

{}From these arguments we estimate the true number of $C_{\rm ACO}\geq
50$ clusters in the near half of the volume to be between 74 and 84,
or 79 $\pm$ 5 (which then implies a total number of 158 $\pm$ 10 such
clusters out to $z = 0.1$ assuming a constant space density). The 79
$\pm$ 5 $C_{\rm ACO}\geq 50$ clusters represent a space density of
$8.6\,\pm\,0.6\,\times10^{-6}\,h^3$ Mpc$^{-3}$. This is slightly
higher than most previous estimates of the density of $R_{\rm ACO}\geq
1$ clusters (e.g.\ by Bahcall \& Soneira 1983, Postman et al.\ 1992,
Peacock \& West 1992, and Zabludoff et al.\ 1993, hereafter ZGHR). The
difference with other work is largely due to our correction of the
intrinsic incompleteness of the ACO catalogue on the basis of the
comparison with the EDCC. Note that our value is quite a bit lower
than that obtained by Scaramella et al.\ (1991) for the Southern ACO
clusters.  These authors found a density of $12.5\times10^{-6}\,h^3$
Mpc$^{-3}$, which does not seem to be consistent with our data.

In the determination of the distribution of velocity dispersions for
the $R_{\rm ACO} \geq 1$ clusters (in Section~6), we will assume that
the incompleteness of the $R_{\rm ACO} \geq 1$ sample can be
corrected for simply by adjusting the density of clusters by the
factor 158/80 (as we have velocity dispersions for only 80 out of an
implied total of 158 clusters). This means that we will assume that
the incompleteness only affects the number of clusters, and that our
estimate of the $\sigma_V$ distribution for $0.08 \la z \leq 0.1$ is
not biased with respect to that found for $z \la 0.08$.

Our assumption that the total number of $R_{\rm ACO} \geq 1$ clusters
is 158 $\pm$ 10 immediately implies that we have missed between 20
and 30 clusters in the outer half of our volume. It is not easy to
account for this number unambiguously from first principles. However,
the number does not seem implausible. Earlier, we estimated from a
comparison with the EDCC that between 10 and 20 clusters have
probably been missed by Abell et al. (1989) in the outer half of the volume
(for whatever reason). This leaves between about 10 and 20 clusters
to be accounted for by two effects: namely the $m_{10}$ limit that we
imposed in the definition of the sample, and the effects of Galactic
obscuration.

Galactic obscuration may indeed have caused some clusters at low
latitudes and close to the redshift limit to have been left out of
the sample. Note, however, that Peacock \& West (1992) argue quite
convincingly that the effects of Galactic obscuration do not operate
below $z\approx 0.08$, a conclusion that seems well supported by the
data in Fig.~3. For $R\geq 1$ clusters at latitudes $|b|\geq 30\degr$
and with distance class $D \leq 4$ (i.e.\ $m_{10}\la 16.4$) Bahcall
\& Soneira (1983) and Postman et al.\ (1992) propose a
cluster selection function varying with galactic latitude as
$$ P(b)=10^{0.32(1-\csc|b|)}. $$ This function also seems to give an
acceptable description for Southern clusters with distance class 5 and
6, and is supposed to be largely caused by the effects of Galactic
obscuration. For our sample, this would imply that we have missed
about 13\% of the estimated total of 158, or about 21 clusters, as a
result of Galactic obscuration. In the light of the result of Peacock
\& West (1992), as well as the data in Fig.~3, all these missing
clusters must have $z
\ga 0.08$.

Within the uncertainties, our interpretation of the observed redshift
distribution of $R_{\rm ACO} \geq 1$ clusters thus seems to be
consistent with all available information.

\subsection{The Sample of 33 Clusters with $C_{\rm 3D} \ge 75$ and $N
            \ge 10$}

In Section~3 we argued, on the basis of the data in Fig.~1$c$, that
our cluster sample with $C_{\rm ACO} \ge 50$ probably is an
acceptable substitute for a complete sample with $C_{\rm 3D} \ge 50$.
The reason for this is that the two richness values scatter around
the $C_{\rm ACO} = C_{\rm 3D}$-line, while the scatter (even though
it is appreciable) appears quite symmetric around this line. From the
same Figure it is also clear that it is not practically feasible to
construct a sample complete down $C_{\rm 3D} = 50$ on the basis of
the ACO catalogue. That would require the ACO catalogue to be
complete down to $C_{\rm ACO} \approx 20$, given the width of the
$f_{\rm main}$ distribution.

However, the data in Fig.~1$c$ also show that from our $C_{\rm ACO}
\ge 50$ complete sample it is possible to construct a subsample that
is complete with respect to intrinsic richness for $C_{\rm 3D} \ge
75$. In Tab.~1$a$ there are 34 clusters with $C_{3D} \ge 75$, for one of
which no velocity dispersion could be determined. In addition, there
are 33 clusters in Tab.~1$a$ that have $C_{\rm ACO} + C_{\rm bck}>75.0$
but for which $f_{\rm main}$ is not available. Using the
distribution of $f_{\rm main}$ given in Fig.~1$a$, we estimate that
10.2 of these would turn out to have $C_{\rm 3D} \ge 75$ if we
measured their $f_{\rm main}$. This brings the estimated total number
of clusters with $C_{\rm 3D} \ge 75$ in Tab.~1$a$ to 44.2.

Finally, one must add an estimated contribution to this sample of
$C_{\rm 3D}\ga 75$ clusters that have probably been missed by Abell et al.\
(1989). As before, the comparison between ACO and EDCC allows us to
estimate this contribution. In principle, one would want to estimate
the number of clusters missed with $C_{\rm 3D} \ge 75$. As the
richnesses of 2 of the 5 clusters missed by ACO in the solid angle of
the EDCC are uncertain, this cannot be done. Therefore, we will
assume that the distribution over richness of the 5 clusters missed
is the same as that of all clusters in our sample. Hence, as 44.2 of
the 128 clusters with $C_{\rm ACO}$ have $C_{\rm 3D} \ge 75$, we
estimate that 1.8 of the 5 missing clusters have $C_{\rm 3D} \ge 75$.
Taking into account the ratio of the solid angles (see Section~4.1)
this implies that 9.2 clusters with $C_{\rm 3D} \ge 75$ have been
missed by ACO in our `cone' volume. This brings the estimated total
number of such clusters in the `cone' to $53.4 \pm 5$, which represents
a density of $2.9 \pm 0.3 \times10^{-6}\,h^3$ Mpc$^{-3}$.

\subsection{Some Remarks on the Quality of the ACO Catalogue}

Since serious doubts have been raised over the completeness and
reliability of the ACO catalogue, it may be useful to summarize here
our findings about its quality.

As was shown in Paper I, the redshift data from the ENACS show that
almost all $R_{\rm ACO} \geq 1$ cluster candidates with $m_{10}(V)
\leq 16.9$ and $b\leq -30\degr$ correspond to real systems that are
compact in redshift space. In only about 10\% of the cases an $R_{\rm
ACO} \geq 1$ cluster candidate appears to be the result of a
superposition of two almost equally rich (but relatively poorer)
systems.

Comparison between the EDCC and ACO catalogues shows that at most
$\approx$15\% (i.e. $25/158$) of the $C_{\rm EDCC} \geq 30$ clusters
(which are expected to have $C_{\rm ACO} \geq 50$) with $m_{10}(V)
\leq 16.9$ in the EDCC do not appear in the ACO catalogue. From this,
one can conclude that {\em the ACO catalogue is at least 85\%
complete} for $R_{\rm ACO} \ge 1$ clusters out to a redshift $z
\approx 0.1$ (see also Briel \& Henry 1993).
Out to $z = 0.08$ the completeness is even higher, viz.
94\%. If one takes into account the effects of Galactic
obscuration the overall completeness of the ACO catalogue for
$|b|\geq 30\degr$ apparently decreases to about 80\% (viz.
$128/158$).

On average, about three quarters of the galaxies in the direction of
$R_{\rm ACO} \geq 1$ clusters with $z\la 0.1$ are in the main system,
i.e.\ the effect of fore- and background contamination is
substantial. However, if one takes into account the effect of field
contamination by deriving $C_{\rm 3D}$, the intrinsic 3-D richness of
the clusters, it appears that the field-corrected ACO richness is
statistically equivalent to the intrinsic richness. This means that
one can use a complete sample of clusters with $C_{\rm ACO} \ge 50$
to investigate the statistical properties of a sample of clusters
complete down to $C_{\rm 3D} \approx 50$. The relation between $C_{\rm
ACO}$ and $C_{\rm 3D}$ however shows a large spread; as a result it
is not possible to select from the $C_{\rm ACO} \ge 50$ cluster
sample a subsample that is complete with respect to\ $C_{\rm 3D}$ for values
of $C_{\rm 3D}$ less than about 75.

We have thus demonstrated that our sample of 128 clusters in the ACO
catalogue with $C_{\rm ACO} \ge 50$ and $z\le 0.1$ can be used as a
statistical sample for the study of the properties of clusters of
galaxies. The subsample of 33 clusters with $C_{\rm 3D} \ge 75$ and
$N \ge 10$ is truly complete with respect to  $C_{\rm 3D}$ and can
therefore be used to check the results from the larger sample.

\section{The Estimation of the Velocity Dispersions}

For a determination of the distribution of cluster velocity
dispersions, one must address several points. First, it is very
important that the individual estimates of the global velocity
dispersions are as unbiased as possible, as any bias may
systematically alter the shape and amplitude of the distribution. For
example, when we identified the systems in velocity space using a
fixed velocity gap, we did not discuss the plausibility of membership
of individual galaxies. Before calculating the global velocity
dispersion we must take special care to remove fore- and background
galaxies that cannot be members of the system on physical grounds.
Leaving such non-members in the system will in general lead to an
overestimation of the global velocity dispersion. Secondly, it has
been shown that the velocity dispersion may vary with position in the
cluster, so that the global velocity dispersion can depend on the
size of the aperture within which it is calculated. Finally, radial
velocities are generally measured only for a bright subset of the
galaxy population. If luminosity segregation is present this will
generally cause the velocity dispersion to be underestimated.

\subsection{The Removal of Interloper Galaxies}

It is well-known that in determining velocity dispersions one has to
be very careful not to overestimate $\sigma_V$ as a result of the
presence of non-members or `outliers'. Recently, den~Hartog \&
Katgert (1995) have shown that velocity dispersions will be
overestimated due to the presence of `interlopers', i.e.\ due to
galaxies that have `survived' the 1000 km/s fixed-gap criterion for
membership, but that are nevertheless unlikely to be members of the
cluster. For the removal of such interlopers, these authors developed
an iterative procedure that employs the combined positional and
velocity information to identify galaxies that are probably not
cluster members.

The procedure starts by estimating a mass profile from an application
of the virial theorem to galaxies in concentric (cylindrical)
cross-sections through the cluster with varying radii. Subsequently,
for each individual galaxy one investigates whether the observed
line-of-sight velocity is consistent with the galaxy being on a
radial orbit with a velocity less than the escape velocity, or on a
bound circular orbit.  If the observed velocity is inconsistent with
either of these extreme assumptions about the shape of the orbit, the
galaxy is flagged as an interloper (i.e.\ a non-member), and not used
in the computation of the mass profile in the next iteration step.
This procedure is repeated until the number of member galaxies
becomes stable, which usually happens after only a few iteration
steps.

In order to ensure that the definition of an interloper and the value
of the cluster velocity dispersion is independent of the redshift of
the cluster, it is necessary to convert the velocities of galaxies
with respect to the cluster to the rest frame of the cluster (e.g.\
Danese et al.\ 1980). Because the elimination of interlopers changes
the estimated average cluster redshift (but only slightly) this
correction is applied to the original data in each iteration step.

\begin{figure}[htb]
\vbox{}
\caption[]{
 The decrease in velocity dispersion as a result of the removal
 of interlopers, versus the initial, uncorrected
 value of the velocity dispersion.}
\label{f:intlop}
\end{figure}

The procedure has been tested on the set of 75 model clusters
presented by Van Kampen (1994). This set of model clusters is
designed to mimic a sample complete with respect to total mass in a
volume that is comparable to that of our $z \le 0.1$ sample. The
initial conditions were generated for an $\Omega=1$ CDM scenario.
Individual cluster models have reasonable mass resolution and contain
dark matter particles as well as soft galaxy particles that are
formed according to a prescription that involves percolation and a
virial condition. A typical simulation has a volume of about
$(30$\Mpc$)^3$ and contains $O(10^5)$ particles. In these models the
membership of galaxies follows unequivocally from the position with
respect to the turn-around radius. It appears that the interloper
removal works very well: in the central region (i.e.\ within the
Abell radius) 90\% of the non-members are indeed removed, and those
that are not removed have a velocity dispersion that is essentialy
equal to that of the member galaxies. In the same region only 0.4\%
of the cluster members is inadvertently removed.

Because the procedure by which we removed interlopers requires a
reliable position for the cluster centre and a reasonably
well-determined mass profile, we have applied it only to the 28
clusters for which at least 50 redshifts are available. This means
that the density of systems with the largest velocity dispersions may
still be somewhat overestimated, as some of the largest dispersions
are for systems with less than 50 redshifts. As a result the
dispersion distribution could in reality fall off even slightly
steeper towards high dispersions at the high end than it appears to
do. However, for the clusters with less than 50 redshifts we have
used the robust biweight estimator for the velocity dispersion (see
Beers et al.\,1990), so that the influence of unremoved interlopers
in the tails of the velocity distribution is strongly reduced.

We have used as many redshifts as possible for each cluster. In 5
cases (i.e. for A0119, A0151, A2717, A3128, A3667) we have combined
existing data from the literature with the new redshifts obtained in
the ENACS. Before combining the two sets of data we have investigated
the consistency of the two redshift scales. The comparison is made
for galaxies of which the position is the same in both surveys to
within 20\arcsec. The redshift scales generally agree to within the
uncertainties (see also Tab.~2 in Paper I).

In Fig.~4 we show, for the 28 systems with at least 50 members, the
decrease of the global velocity dispersion as a function of the value
of the dispersion before the interlopers were removed. For
dispersions below about 900 km/s the reduction is at most about 10\%.
It is clear that the decrease can be much larger for the largest
dispersions, with reductions of as much as 25 to 30\%.  The point
near the upper right-hand corner refers to A151, before and after
the separation of the 2 low redshift systems.  In two clusters,
A85 and A3151, the interloper removal failed to delete a group of
interlopers. In the wedge diagrams the two groups were clearly
compact in velocity and spatial extent, with a velocity offset of
$\approx 2000$ km/s with respect to the main system. We have removed
these groups by hand.

Our analysis indicates that systems with global velocity dispersions
larger than 1200 km/s have such low space density (if they exist at
all) that in our volume they do not occur. This would seem to be at
variance with the result of ZGHR, who find one cluster, A2152, with a
dispersion of 1346 km/s in a sample of 25 clusters, in a volume that
is about a factor of 5 smaller than ours. However, it must be said
that, had the cluster A2152 appeared in our sample (with the same
number of redshifts that ZGHR had available, viz. 21), we would probably
also have found a high velocity dispersion. However, had 50 redshifts
been available, we would almost certainly have eliminated quite a few
interlopers, and would thereby probably have reduced the dispersion
substantially. Therefore, a distribution of
velocity dispersions becomes less biased towards high dispersion
values, when the average number of redshifts per cluster increases.

The absence of systems with very large velocity dispersions in our
sample is due to the large fraction of systems for which we could
eliminate non-members using a physical criterion. As we have discussed
extensively in Paper I, it is very unlikely that the absence of very
large  velocity dispersions in our sample is due to the method by which
we defined the systems in the first place. If we had used the method
of ZGHR to define the systems (see Tab.~6 in Paper I for detailed
information) one or two systems which we have broken up would have
remained single. However, when we identify our clusters with the ZGHR
method and subsequently remove the interlopers, we find essentially
the same clusters with the same velocity dispersions. It thus appears
that quite a few of the galaxies in the clusters for which ZGHR find
a very large velocity dispersion are unlikely to be members of the
system.

\subsection{The Effects of Aperture Variations and Luminosity
            Segregation}

Another possible bias in estimating velocity dispersions is due to
the fact that the velocity dispersion frequently varies significantly
with distance from the cluster centre (see e.g.\ den~Hartog and
Katgert, 1995). Differences in the physical size of the aperture
within which the velocities are measured are inevitable for a sample
of clusters with redshifts between 0.02 and 0.1. However, the sizes
of the apertures used in the observations have a dispersion of only
about 30\% around the average value of about 0.9 \Mpc. On the basis
of the velocity dispersion profiles discussed by den~Hartog and
Katgert (ibid.) and in agreement with Girardi et al.\ (1993) we
estimate that corrections for variations in the aperture in practice
are at most about 10\% (and can be both positive and negative). They
are thus substantially smaller than the largest corrections applied
to the dispersions due to interloper removal (which are exclusively
negative). We have decided not to attempt to apply corrections for
aperture variations based on an average velocity dispersion profile,
as this will only introduce noise and is not expected to change the
results in a systematic way.

den~Hartog and Katgert (ibid.) found signs of luminosity segregation in
approximately 20\% of the clusters in their sample. Hence, it is
necessary to check that our velocity dispersions are not biased in a
systematic way as a result of the fact that we have sampled for many
clusters only the brightest galaxies in the central regions of the
clusters. Luminosity segregation is a manifestation of the physical
processes of mass segregation (heavy galaxies move slower and their
distribution is more centrally concentrated than that of light
galaxies) and velocity bias (the velocity dispersion of the galaxies
is lower than that of the dark matter particles). The reality of mass
segregation and velocity bias is still a matter of dispute (see e.g.\
Carlberg 1994, versus e.g.\ Katz et al.\ 1992, Biviano et al.\ 1992,
Lubin \& Bahcall 1993, and Van Kampen 1995). Moreover, significant
mass segregation may not be readily observable if it is accompanied
by significant variations in \MLR.

We have tested for the presence of luminosity segregation in our
cluster sample, and will discuss the results in a forthcoming paper.
It appears that luminosity segregation exists, but that it is
exclusively linked to the very brightest cluster galaxies, which
appear to move very much slower than the other galaxies. As our
velocity dispersions are based on at least 10 (but often many more)
redshifts, the effect of luminosity segregation is completely
negligible in the context of the present discussion.

\section{The Distribution of Velocity Dispersions}

By virtue of the significant (but very broad) correlation between
richness and velocity dispersion, the largest velocity dispersions
are in general found in the systems with the highest richnesses,
whereas the low dispersions are found preferentially in the poorer
systems. As a result, any sample of clusters presently available is
biased against low velocity dispersions, because of the lower limit
in richness that defines the sample. This means that any observed
(and predicted!) distribution of cluster velocity dispersions refers
to a specific richness limit. As a matter of fact, the distribution
will in principle be biased for velocity dispersions smaller than the
largest value found at the richness limit; above the latter value the
distribution is unbiased.

\begin{figure}[t]
\vbox{}
\caption[]{
The apparent distribution of cluster velocity dispersions,
$a)$ for the sample of 80 clusters with $C_{\rm ACO}\geq 50$ and
$z\leq 0.1$; $b)$ for the subsample of 33 clusters with
$C_{\rm 3D}\geq 75$.}
\label{f:histo}
\end{figure}

Our estimates of the apparent distributions of $\sigma_V$ are shown
in Figs.~5$a$ and $b$. They refer to the two subsets of clusters:
viz.\ the complete sample of 80 clusters with $z\leq 0.1$ and $C_{\rm
ACO}\geq 50$ and the complete subsample of 33 clusters with $C_{\rm
3D}\geq 75$. As we discussed before, the latter is truly complete
with respect to intrinsic 3-D richness (i.e.\ fore- and background
contamination has been taken into account) and the completeness with
respect to redshift is beyond suspicion. We also found that the
sample of 80 $C_{\rm ACO}\geq 50$ clusters can be used as a
substitute for a $C_{\rm 3D} \ge 50$ sample. However, as it is somewhat
incomplete near the redshift limit we can derive an unbiased estimate
of the $\sigma_V$ distribution only if there is no systematic change
in our sample of $\sigma_V$ with redshift.

In Fig.~6 we show that there is indeed no evidence for a significant
correlation of velocity dispersion with redshift in our sample. The
lack of systems with $\sigma_V \ga$ 900 km/s below a redshift of
about 0.05 is not considered significant in view of the small volume
sampled. It is also encouraging that there is no clear bias against
systems with low values of $\sigma_V$, between $z=0.08$ and $z=0.1$.
Therefore, the data in Fig.~6 indicate that the result for the sample
of 80 systems with $z\leq 0.1$ and with $C_{\rm ACO}\geq 50$ should
be equally reliable as that for the subsample of 33 $C_{\rm 3D}\geq
75$ systems, with the advantage of larger statistical weight.

\begin{figure}[t]
\vbox{}
\caption[]{
 Distribution with respect to redshift and velocity dispersion of the
 80 systems with $C_{\rm ACO}\geq 50$ and $z\le 0.1$.}
\label{f:svvsz}
\end{figure}

In Fig.~7 we show the cumulative distribution of $\sigma_V$ for the
sample of 80 clusters with $C_{\rm ACO} \ge 50$ (full-drawn line) and
for the subsample of 33 clusters with $C_{\rm 3D} \ge 75$ (dashed
line). Note that the densities follow directly from the space
densities that we derived in Sections 4.1 and 4.2, and have thus not
been scaled to some external cluster density (as is sometimes done in
the literature). For comparison, and to illustrate the effect of the
interloper removal, we also show the cumulative distribution of
$\sigma_V$ for the 80 cluster sample, when no interlopers are
removed, i.e\ with $\sigma_V$ for the clusters with 50 redshifts or
more determined only with the robust biweight estimator (dotted
line). It is clear that without interloper removal the distribution
of $\sigma_V$ is significantly biased for $\sigma_V \ga 800$ km/s.

The (interloper-corrected) distributions for the two samples agree
very well for $\sigma_V \ga 900$ km/s. This is not surprising because,
as we will see below, there are hardly any clusters with $\sigma_V \ge
900$ km/s that have $C_{\rm 3D} < 75$. The good agreement therefore
just shows that the ratio of the space densities derived in Sections
4.1 and 4.2 is quite good. The two distributions also illustrate the
bias against low values of $\sigma_V$. The value of $\sigma_V$ at
which the bias sets in and the magnitude of the bias are seen to
depend on the richness completeness limit of the sample in a way that
is consistent with the discussion at the beginning of this
Section. For the sample with $C_{\rm 3D} \ge 75$ the incompleteness
starts at $\sigma_V \approx 900$ km/s, while for the sample with
$C_{\rm ACO} \ge 50$ it starts at $\sigma_V \approx 800$ km/s.

\begin{figure}[t]
\vbox{}
\caption[]{
 The cumulative distribution of cluster velocity dispersions.
 Solid line:  for the sample of 80 clusters with $C_{\rm
              ACO}\geq 50$ and $z\leq 0.1$.
 Dotted line: for the sample of 80 clusters with $C_{\rm
              ACO}\geq 50$ and $z\leq 0.1$, but without interlopers
              removed.
 Dashed line: for the sample of 33 clusters with $C_{\rm
              3D}\geq 75$ and $z\leq 0.1$.}
\label{f:cumdis}
\end{figure}

For $\sigma_V\geq 800$ km/s the cumulative distribution for the
sample of 80 clusters can be parametrized as follows:

$$\log n(>\sigma_V) = -5.6 -0.0036
  (\sigma_V - 800 \mbox{\,km/s}) \quad\mbox{$h^3$ Mpc$^{-3}$} $$

For $\sigma_V < 800$ km/s the same distribution also seems to be
described fairly accurately by a power law, but the significance of
that fit is much less apparent because of the bias that is likely to
increase with decreasing $\sigma_V$.

ZGHR have tried to correct the bias against low-velocity dispersion
systems by combining clusters and dense groups. Indeed, it appears
that continuation of the above power law fit down to $\sigma_V$ = 700
km/s would predict, within the errors, the correct density
$n(>700\,\mbox{km/s})$ for the combination of clusters and dense
groups. However, as the definition and selection of dense groups is
different from that of rich clusters, it is not unlikely that the
intrinsic properties of the groups, such as $\sigma_V$, as well as
their spatial density may differ systematically from that of the
clusters. Also, for the dense groups a similar bias operates as for
the clusters. Combination of the two $\sigma_V$ distributions is
therefore not without problems.

It is of some interest to have a closer
look at the values of $\sigma_V$ below which the two distributions
are biased as a result
of the lower limits in $C_{\rm ACO}$ and $C_{\rm 3D}$ that define the
samples. These values of $\sigma_V$ are the maximum values found
near the cut-off in richness, and they can be estimated from Fig.~8,
in which we show several distributions of $\sigma_V$ against
richness. From the distribution of $\sigma_V$ against $C_{\rm ACO}$,
shown in Fig.~8$a$, it appears that the maximum $\sigma_V$ near the
richness limit $C_{\rm ACO}=50$ is about 800 km/s.

\begin{figure*}
\vbox{}
\caption[]{
Velocity dispersion $\sigma_V$ versus richness.\\
$a)$ $\sigma_V$ vs.\ $C_{\rm ACO}$ for all main systems
in the ENACS sample.\\
$b)$ $\sigma_V$ vs.\ $C_{\rm EDCC}+20$ (squares) and
$\sigma_V$ vs.\ $C_{\rm 2D}$, obtained by Dalton (1992) from
application of the ACO cluster definition to candidate clusters in
the APM survey (diamonds).\\
$c)$ $\sigma_V$ vs.\ $C_{\rm APM}$, from the APM cluster
catalogue listed in Dalton et al.\ (1994).\\
$d)$ As $a)$, but now for $C_{\rm 3D} = f_{\rm main} \times (C_{\rm
ACO} + C_{\rm bck})$, i.e.\ the intrinsic richness corrected for
superposition effects.\\
$e)$ As $d)$, but now for machine-based `ACO' counts corrected for
superposition effects.
Note that the EDCC counts were also corrected, viz.\
$C_{\rm 3D} = f_{\rm main} \times (C_{\rm EDCC}+n_{\rm bck})$, where $n_{\rm
bck}$ is the number of background galaxies estimated by LNCG.\\
$f)$ As $d)$ but now for the APM clusters, corrected for superposition effects.
}
\end{figure*}

In Fig.~8$b$ we show $\sigma_V$ vs.\ richness for the subset of ENACS
clusters for which either LNCG (squares) or Dalton (priv.~comm.;
diamonds) give an alternative, machine-based estimate of the
richness. In Fig.~8$b$ the ordinate is $C_{\rm EDCC}+20$ rather than
$C_{\rm EDCC}$, because there seems to be a systematic offset between
$C_{\rm ACO}$ and $C_{\rm EDCC}$ of about 20 (see Section~4.1).  It is
clear that for a richness limit of 50 in the machine-based counts,
the bias is again absent only for $\sigma_V \ga 800$ km/s.

In Fig.~8$c$ we show velocity dispersion vs. richness count $C_{\rm
APM}$ from the APM cluster catalogue, for the 37 clusters in the APM
catalogue of which the positions coincide with that of a cluster in
our sample to within half an Abell radius. Note that Dalton et al.\
(1994) have calculated the richness inside half an Abell radius and
within a variable magnitude interval based on the luminosity function
in the region of the cluster, in order to be less sensitive to
interlopers. As a result $C_{\rm APM}$ is systematically lower than
$C_{\rm ACO}$, and the richness limit $C_{\rm ACO}=50$ corresponds to
$C_{\rm APM}=35$ (Efstathiou et al.\ 1992a).  From Fig.~8$c$ we
conclude that the bias against
low velocity dispersions sets in at $\sigma_V \approx 800$ km/s.

{}From the fact that the three left-hand panels in Fig.~8 are
qualitatively very similar we conclude that the large spread in
Fig.~8$a$ is not primarily due to errors in the values of $C_{\rm
ACO}$, as a similar spread is seen for the two other catalogues.
Therefore, we conclude that the large spread in velocity dispersion
for a fixed value of 2-D richness is probably (at least
partially) intrinsic to the clusters.

\begin{figure*}[htb]
\vbox{}
\caption[]{
 Comparison of the present cumulative distributions of velocity
 dispersions with distributions from the literature.
 The solid line in both figures refers to our sample of 80 clusters
 with $z\le 0.1$.
 a) Optical data: Girardi et al.\ (1993, dotted line),  ZGHR
    (short dashed line).
 b) X-ray data: Henry \& Arnaud (1991, dashed line) and
    Edge et al.\ (1990, dotted line).}
\end{figure*}

In Fig.~8$d$ we show the relation between  $\sigma_V$ and $C_{\rm
3D}$, with the latter based on $C_{\rm ACO}$. It appears that the
relation is less broad than that in Fig.~8$a$.  Apparently, the
correction for superposition effects (which we could only apply
thanks to the redshift information) results in a fairly significant
decrease of the spread in the relation. In Fig.~8$d$ the existence of
an upper limit to the velocity dispersion of $\approx$900 km/s at the
richness limit of 75 is very clearly illustrated. The spread in the
relation between $\sigma_V$ and $C_{\rm 3D}$ in Fig.~8$d$ is
probably not primarily due to errors in the values $C_{\rm ACO}$.
This is supported by the data in Figs.~8$e$ and 8$f$, where we show
the relation between $\sigma_V$ and machine-based counts that have
been corrected for superposition effects.

We conclude therefore that the scatter between $\sigma_V$ and
richness (in whichever way it is measured) must largely be intrinsic.
In other words: a given velocity dispersion may be found in clusters
of quite different richnesses, while clusters of a given richness
span a large range of velocity dispersion.

\section{Discussion}

In the following we will make two types of comparison of the results
obtained here with earlier results. First we will compare with other
determinations of the cumulative distribution of cluster velocity
dispersions $n(>\sigma_V)$, as well as with the distributions of
cluster X-ray temperatures $n(>T_X)$. Subsequently, we will discuss
the relation between our result and some model predictions for
$n(>\sigma_V)$ from the literature.

\subsection{Comparison with Other Data}

There are several other determinations of $n(>\sigma_V)$ in the
literature. Recent papers on the subject are e.g.\ those by Girardi
et al.\ (1993), and by ZGHR. The result of Girardi et al. (1993) is based
on a compilation of redshifts for cluster galaxies. As a result, the
amplitude of $n(>\sigma_V)$ is not known in absolute terms, but has
been inferred from the integrated fraction of clusters together with
an external estimate of the total density of rich clusters. Collins
et al.\ (1995) also present a distribution of $\sigma_V$ that is not
normalized. On the contrary, ZHGR present, like we do, an estimate of
$n(>\sigma_V)$ with a calibrated space density. A comparison with the
results of Girardi et al.\ (1993) and ZGHR is given in Fig.~9$a$,
where the result of Girardi et al. (1993) has been scaled to the density of
rich clusters derived in Section~4.1, rather than that given by
Bahcall and Soneira (1983).

Although the previous estimates of $n(>\sigma_V)$ involved clipping
of `outliers', none employed the removal of interlopers as described
in Section~5.1 , and therefore it is not too surprising that for
$\sigma_V \ga 900$ km/s our result is systematically lower than the
other two. Girardi et al. (1993) obtain a similar slope but a (perhaps
not very certain) amplitude that is at least two times higher than
ours. We do not show the result of Collins et al.\ separately as it
appears to agree with that of Girardi et al. On the other hand, the
result of ZGHR agrees very nicely with ours for $\sigma_V \la 900$
km/s, but for larger values of $\sigma_V$ they obtain a slope that is
definitely less steep than ours.

Our upper limit on the occurence of clusters with $\sigma_V \ga$ 1200
km/s is much more severe than any previous result based on optical data,
namely that the space density of such clusters is less than one in
our survey volume of $1.8\times10^7\,h^{-3} {\rm Mpc}^3$. As we
discussed above, this is almost entirely due to our removal from the
redshift data of those interlopers that can only be recognized on the
basis of the combination of radial velocity {\em and} projected
position within the cluster.

In Fig.~9$b$ we compare our result with distributions of the cluster
X-ray temperature $T_X$ by Henry \& Arnaud (1991) and by Edge et al.
(1990). In transforming the $T_X$ scale into a $\sigma_V$ scale we
assumed that $\sigma_V^2  = (kT_X/ \mu m_H)$, where $\mu$ and $m_H$
have their usual meaning. The reason for the discrepancy between the
two X-ray results is not known. ZGHR have suggested that the
discrepancy is due to differences in normalization caused by
different fitting procedures, sample size and sample completeness.
The agreement between our result and that of Henry \& Arnaud (1991) is
excellent for $\sigma_V \ga 800$ km/s. Both the amplitude and the
slope agree very well, and to us this suggests that the removal of
interlopers is necessary, and that our removal procedure is adequate.
It also suggests that the velocity dispersions in excess of 1200
km/s, found by others, must indeed almost all be overestimates caused
by interlopers. Interestingly, the two results start to diverge below
$\approx$800 km/s. Although one cannot claim that $n(>T_X)$ is very
well determined in that range there is at least no contradiction with
the conclusion that we reached in Section~6, namely that our
$n(>\sigma_V)$ must start to become underestimated below $\approx$800
km/s as a result of the richness limit of our cluster sample.

The extremely good agreement between our $n(>\sigma_V)$ and the
$n(>T_X)$ by Henry \& Arnaud (1991) for $\sigma_V \ga$ 800 km/s, for an
assumed value of $\beta = \sigma_V^2 /(kT_X / \mu m_H) = 1$ strongly
suggests that X-ray temperatures and velocity dispersions
statistically measure the same cluster property. This is in agreement
with earlier results of Lubin \& Bahcall (1993), Gerbal et al.\ (1994), and
den~Hartog \& Katgert (1995) who also find that it is
not necessary that the ratio of `dynamical' and X-ray temperatures
differs from 1.0. Of course, in our case this statement only refers
to the {\em sample} of clusters, and it has not been proven to be
valid for individual clusters. On the basis of the data in Fig.~9$b$
we conclude that the {\em average} value of $\beta$ must lie between
0.7 (the value required if the upper range of our $n(>\sigma_V)$
determinations must coincide with the result of Edge et al. 1990), and
1.1 (the value required if the lower range of our data must coincide
with the result of Henry \& Arnaud 1991).

\subsection{Requirements for Useful Comparison with Models}

There are quite a few papers in the literature in which model
calculations of clusters of galaxies are presented from which one
can, in principle, derive model predictions of $n(>\sigma_V)$. These
models are generally of two kinds. First, there are numerical (or
analytical) models of a sufficiently large cosmological volume,
containing a sample of clusters, each of which is modeled with
relatively low resolution. In this case one can obtain a direct
estimate of $n(>\sigma_V)$, the normalization of which is
unambiguous. A good example of this type of model was described by
FWED. Secondly, sets of higher-resolution
cluster simulations may be created for which the global properties
are distributed as predicted for an arbitrarily chosen, large
cosmological volume. In this case, the normalization of
$n(>\sigma_V)$ depends on the details of the selection of the set of
cluster models. An example of the latter has been described by Van
Kampen (1994). Of course, in both cases, the resulting predictions
are valid only for the chosen scenario of large-scale structure
formation. We will limit ourselves here to a brief discussion of
various aspects of the comparison between observations and models,
and demonstrate the use of our result in a comparison with the models
of FWED and Van Kampen (1994).

\begin{figure*}[htb]
\vbox{}
\caption[]{
Comparison of the cumulative distribution of $\sigma_V$ derived in
this paper (solid line) with predictions from
standard CDM N-body simulations with different values of the bias
parameter $b$.
\newline
$a.)$ The dashed lines correspond to distributions of the line-of-sight
$\sigma_V$ computed by FWED in {\em spheres} with radius equal to the Abell
radius. These curves were corrected for the aperture effect, the projection
effect and the softening effect.
The values of $b$ are 2.0, 2.5 and 3.3 (top to bottom).
\newline
$b.)$ Distributions of the line-of-sight
$\sigma_V$ of galaxies in a cylinder with a radius of 1.0 \Mpc  in
Van Kampen's (1994) models. The dashed lines are the distributions for
galaxies, for values of $b$ of 1.6, 2.2, and 2.8 (top to bottom).
The dotted line gives the distribution of $\sigma_V$ computed for the
dark matter inside a cylinder for $b$ = 2.2.
}
\end{figure*}

A meaningful comparison between observations and models requires that
one derives from the models a prediction of exactly the same quantity
as one has observed. As we discussed above, our $\sigma_V$ estimates
refer to a {\em cylinder} with an average radius of 1.0 \Mpc and a
depth of (about) twice the turn-around radius of the cluster. From
their models, FWED have calculated the line-of-sight velocity
dispersion within a {\em sphere} with radius equal to the Abell
radius. As the latter excludes the, mostly slowly moving, galaxies
that are near the turn-around radius, the $\sigma_V$ values in a
sphere are expected to be systematically higher than in a cylinder
with the same radius, by as much as 10 \%. On the other hand, the
value of $\sigma_V$ also depends on the radius of the cylinder or
sphere. On average, $\sigma_V$ is expected to decrease with
increasing radius of the cylinder because, on average, the velocity
dispersion tends to decrease with distance from the cluster centre.
In a comparison between our data and the model prediction of FWED
(who use a sphere with radius 1.5 \Mpc as compared to our cylinder
with radius 1.0 \Mpc) we will assume that the two effects compensate
almost exactly.  We conclude this from a direct comparison between
the two quantities based on the models of Van Kampen (1995). As the models
by FWED and Van Kampen use the same $\Omega=1$ CDM formation
scenario, we assume that this conclusion is also valid for the FWED
models.

The values of the global $\sigma_V$ may depend fairly strongly on the
details of the integration scheme in the N-body simulations.  For
instance, the models of FWED do not have much resolution on the scale
of galaxies, since huge volumes (of the order of the volume of the
ENACS) had to be simulated with $O(2\cdot10^5)$ particles. As a
result, the scale-length for force softening is well over 100 kpc.
Van Kampen (1995) has studied the effect of the softening scale-length on
$\sigma_V$ and finds that for the FWED scale-length the velocity
dispersions are 15--20\% smaller than for a softening-length of 20
kpc.

Another aspect of the comparison is the identification of the
clusters in (particularly) the large-scale simulations. FWED
identified `galaxies', at the end of the simulation, as peaks in the
density field, without altering the dynamical properties of the
constituent dark particles. First, it is not clear whether galaxies
form solely or preferentially from peaks in the initial density field
(see e.g.\ Van de Weygaert \& Babul 1994, and Katz et al.\ 1994).
Secondly, Van Kampen (1995) found that the spatial distribution of
the `galaxies' in his models can differ substantially from that of
the dark matter. The galaxy identification `recipe' can thus
influence the definition of the clusters and of the cluster sample,
as a cluster is identified through the number of galaxies inside an
Abell radius.

Finally, it is possible that in clusters the velocity dispersion of
the galaxies is 10 $-$ 20\% lower than that of the dark matter, as a
result of velocity bias (see e.g.\ Carlberg 1994 and Summers 1993).
The reality of velocity bias is still controversial (see e.g.\ Katz
et al.\ 1992, Lubin \& Bahcall 1993 and Van Kampen 1995), and one
must be careful to derive the velocity dispersion of the {\em galaxies}
from the models.

\subsection{Comparison with Selected Model Predictions}

In Fig.~10 we compare our estimate of $n(>\sigma_V)$ to the model
predictions from FWED and Van Kampen (1994), which both assume an $\Omega=1$
CDM formation scenario. In Fig.~10$a$ we compare our result with the
predictions by FWED, who identified the $R\geq 1$ clusters in their
models as groups of dark and luminous particles for which the
luminosity inside a sphere with Abell radius exceeds 42 $L^*$. We
corrected the FWED velocity dispersions for the effects of the fairly
large softening parameter by multiplying them by a factor of 1.18. We
assumed that the differences related to the use of spherical and
cylindrical volumes, as well as different sizes of the aperture,
compensate. For a suitable choice of the bias parameter the
observations and predictions can be made to agree fairly well,
although one could argue that for $\sigma_V \ga$ 800 km/s the slope
of the observed $n(>\sigma_V)$ is steeper than that of the predicted
$n(>\sigma_V)$ for any bias parameter in the range from 2.0 to 2.5.
This may be (partly) due to the fact that FWED convolved their result
with assumed errors in $\sigma_V$ of about 20 \%, which is probably a
factor of two larger than the errors in our $\sigma_V$ estimates for
$\sigma_V\ga 800$ km/s.

In Fig.~10$b$ we make the comparison with the predictions by Van
Kampen (1994), who applied a $C_{\rm 3D}$ lower limit for identifying the
clusters to be included in the comparison.
We have scaled his results to the density of rich clusters derived
in Section~4.1. For his models we show the
distributions of $\sigma_V$ for the galaxies, but for $b = 2.2$ we
also show the distribution for the dark matter; it is clear that the
galaxies and dark matter give essentially the same $n(>\sigma_V)$.
FWED and Van Kampen (1994) seem to predict different amplitudes of
$n(>\sigma_V)$ for the same values of the bias parameter. We do not
consider this the proper place to investigate possible explanations
for the difference. Suffice it to say that the difference between the
cluster identification schemes may well be one of the causes.  The
observations and predictions can be made to agree fairly well,
although one could again argue that for $\sigma_V \ga$ 800 km/s the
slope of the observed $n(>\sigma_V)$ is significantly steeper than
that of the predicted $n(>\sigma_V)$.

{}From both comparisons we see that for the standard $\Omega=1$ CDM
model a large bias parameter is indicated (between 2.0 and 2.5 for
the FWED models and between 2.4 and 2.8 for the models by Van Kampen 1994).
For the commonly accepted low value of the bias parameter of about
1.0, the models clearly predict too many clusters with large
velocity dispersions. Also, the relative proportions of high- and
low-$\sigma_V$ clusters do not seem to be right.

Our result confirms the conclusions by FWED and White et al.\ (1993)
that the distributions of the velocity dispersions or masses of rich
clusters do not support $\Omega$ = 1 CDM models with low values of the
bias parameter.  The high values of the bias parameter, that one
infers from the comparisons in Fig.~10, are in conflict with the
results for the normalization of the $\Omega=1$ CDM models on larger
scales, from comparisons with e.g.\ the COBE data (Wright et al.\
1992, Efstathiou et al.\ 1992b), the power spectrum analysis of the
QDOT survey (Feldman et al.\ 1994) and the recent analyses of
large-scale streaming (Seljak \& Bertschinger 1994).

The important conclusion is therefore that, for $\sigma_V \ga 800$ km/s
our observed distribution $n(>\sigma_V)$ provides a very powerful
constraint for cosmological scenarios of structure formation. It will
not be too long before detailed predictions based on the currently
fashionable (or other) alternative scenarios (be it low-density,
tilted-spectrum, vacuum-dominated or neutrino-enriched CDM) can be
compared, in a proper way, to the observational constraints. Even
though it is worthwhile to try and obtain unbiased estimates of
$n(>\sigma_V)$ for $\sigma_v \la $800 km/s, it would seem that the
high$-\sigma_V$ tail of the distribution has the largest
discriminating power.

\section{Summary and Conclusions}

We have obtained a statistically reliable distribution of velocity
dispersions which, for $\sigma_V \ga 800$ km/s, is free from biases and
systematic errors, while below 800 km/s it is biased against low
values of $\sigma_V$ in a way that is dictated by the richness limit
of our sample, viz.\ $C_{\rm ACO} \geq 50$.

The observed distribution $n(>\sigma_V)$ offers a reliable constraint
for cosmological scenarios, provided model predictions are based on
line-of-sight velocity dispersions for all galaxies inside the
turn-around radius and inside a projected aperture of 1.0 \Mpc, and
provided the clusters are selected according to a richness limit that
mimics the limit that defines the observed cluster sample.

The sample of ACO clusters with $|b|>30\degr$, $C_{\rm ACO}\geq 50$
and $z\leq 0.1$ is $\approx$85\% complete. We find that the density
of clusters with an apparent richness $C_{\rm ACO} \geq 50$ is $8.6
\pm 0.6 \times 10^{-6}\,h^3$ Mpc$^{-3}$, which is
slightly higher than
earlier determinations (e.g. by Bahcall \& Soneira 1983, Peacock and
West 1992, and ZGHR). We show that one can define a
complete subsample of the $C_{\rm ACO} \geq 50$ sample that contains
all clusters with an intrinsic 3-D richness $C_{\rm 3D} \ge 75$; the
density of the latter is $2.9 \pm 0.3 \times10^{-6}\,h^3$ Mpc$^{-3}$.

We find that cluster richness is a bad predictor of the velocity
dispersion (whether it is based on ACO or machine counts) due to the
very broad correlation between the two cluster properties. It appears
that the spread in this correlation must be largely intrinsic, i.e.
not due to measurement errors. As a result, all samples of clusters
that are selected to be complete with respect to richness are biased
against low-$\sigma_V$ systems.

The space density of clusters with $\sigma_{V}>1200$
km/s is less than $0.54\times10^{-7}\,h^3\,{\rm Mpc}^{-3}$.  This is
in accordance with the limits from the space density of hot X-ray
clusters. From the good agreement between $n(>\sigma_V)$ and
$n(>T_X)$ we conclude that $\beta = \sigma_V^2 /(kT_X / \mu m_H)
\approx 1$ and that X-ray temperature and velocity dispersion are
statistically measuring the same cluster property.

For the low values of the bias parameter ($b\approx 1.0$) that are
implied by the large-scale normalization of the standard $\Omega=1$
CDM scenario for structure formation this model appears to predict
too many clusters with high velocity dispersions. Approximate
agreement between observations and the $\Omega=1$ CDM model can be
obtained for bias parameters in the range $2\la b \la 3$, in
agreement with the earlier conclusions by FWED or White et al.\
(1993).

\begin{acknowledgements}
{We thank Eelco van Kampen for helpful discussions and for allowing
us to use his unpublished cluster models. Gavin Dalton is gratefully
acknowledged for making available a machine-readable version of his
Ph.D.\ thesis, as well as four alternative cluster catalogues based
on the APM galaxy catalogue. We thank Mike West for providing us with
his compilation of cluster redshifts.
We thank the referee, L.\ Guzzo, for several useful comments and for
pointing out an error in the manuscript, the correction of which has
led to an important improvement of the paper. The cooperation between
the members of the project was financially supported by the following
organizations: INSU, GR Cosmologie, Univ. de Provence, Univ. de
Montpellier (France), CNRS-NWO (France and the Netherlands), Leiden
Observatory, Leids Kerkhoven-Bosscha Fonds (the Netherlands),
Univ. of Bologna, Univ.
of Trieste (Italy), the Swiss National Science Foundation, the
Ministerio de Educacion y Ciencia (Spain), CNRS-CSIC (France and
Spain) and by the EC HCM programme.}
\end{acknowledgements}

\end{document}